\documentclass[a4paper,preprint]{emulateapj}
\usepackage{amstext}
\usepackage{apjfonts}
\usepackage{amsmath}

\shorttitle{Stability of Magnetic Wind-Driving Disks}
\shortauthors{K\"onigl}
\submitted{\quad}
\begin{document}


\title{Are Magnetic Wind-Driving Disks Inherently Unstable?}


\author{Arieh K\"onigl}
\affil{Department of Astronomy \& Astrophysics and Enrico Fermi
Institute,\\ University of Chicago, 5640 S. Ellis Avenue, Chicago, IL 60637}
\email{arieh@jets.uchicago.edu}

%\and

%\author{R. J. Hanisch\altaffilmark{5}}
%\affil{Space Telescope Science Institute, Baltimore, MD 21218}



\begin{abstract}
There have been claims in the literature that accretion disks in
which a centrifugally driven wind is the dominant mode of
angular momentum transport are inherently unstable. This issue
is considered here by applying an equilibrium-curve
analysis to the wind-driving, ambipolar diffusion-dominated,
magnetic disk model of \citet{WK93}. The
equilibrium solution curves for this class of models typically
exhibit two distinct branches. It is argued that only one of
these branches represents unstable equilibria and that a real
disk/wind system likely corresponds to a stable solution.
\end{abstract}


%% subject header.  Objects should be in the appropriate "individual"
%% headers (e.g. quasars: individual, stars: individual, etc.) with the
%% additional provision that the total number of headers, including each
%% individual object, not exceed six.  The \objectname{} macro, and its
%% alias \object{}, is used to mark each object.  The macro takes the object
%% name as its primary argument.  This name will appear in the paper
%% and serve as the link's anchor in the electronic edition if the name
%% is recognized by the data centers.  The macro also takes an optional
%% argument in parentheses in cases where the data center identification
%% differs from what is to be printed in the paper.


%\keywords{globular clusters: general ---
%globular clusters: individual(\objectname{NGC 6397},
%\object{NGC 6624}, \objectname[M 15]{NGC 7078},
%\object[Cl 1938-341]{Terzan 8})}


\section{Introduction}
\label{introduction}

The ubiquity of energetic and highly collimated jets in compact
astronomical objects (ranging from young stellar objects to
active galactic nuclei) is often interpreted in terms of a
universal mechanism: hydromagnetic outflows from accretion disks
\citep[e.g.,][]{L00}. In a pioneering paper that presented
semianalytic self-similar solutions for centrifugally driven outflows,
\citet{BP82} demonstrated that such jets could efficiently
transport angular momentum from the underlying disks and
suggested this as an alternative mechanism to the radial viscous
angular-momentum transport that has traditionally been invoked
in accretion-disk models. \citet{K89} and \citeauthor{WK93}
(\citeyear{WK93}, hereafter WK93) subsequently incorporated
the disk into the self-similar wind model and suggested
that the apparent connection between disks and jets may at least in part be a
reflection of the fact that the vertical magnetic
angular-momentum transport is a necessary ingredient in the
accretion process. Additional semianalytic models
have since been constructed by several authors
\citep[e.g.,][]{FP93a,FP93b,FP95,Li95,Li96,F97,OL98,OL01,CF00a,CF00b,FC04}
and have served to refine our
understanding of the equilibrium configurations. However, the
stability of the derived disk/wind solutions is still being debated.

WK93 and \citet{KW96} argued that wind-driving disks
should be immune to the most powerful cataloged
disk instabilities. In particular, they pointed out that disks in
which a centrifugally driven outflow transports all the
liberated angular momentum naturally lie in a stability
``window'' in which the magnetic field is strong enough not to
be affected by the magnetorotational instability
\citep[e.g.,][]{BH98} but is not so strong as to be subject to a
radial interchange instability \citep[e.g.,][]{STP95}. However,
\citeauthor{LPP94} (\citeyear{LPP94}, hereafter LPP94) 
suggested that magnetic wind-driving disks may nevertheless be inherently
unstable. Based on a simplified model, they derived two
relations between the mass outflow rate per unit
area and the infall speed at a given disk radius: one of these relations
yielded an S-shaped curve and the other a monotonically
increasing curve. LPP94 argued that disks that lose
angular momentum smoothly at all radii through a wind that
removes only a small fraction of the inflowing mass correspond
to the two curves intersecting in the middle portion of the S
curve and are therefore unstable. This derivation was
questioned by \citet{KW96}, who argued that the equilibrium
model adopted by LPP94 was overly simplified. In particular, they applied the
LPP94 prescription to the more comprehensive model considered by WK93
and showed that both of the resulting equations for the mass
outflow rate were independent of the inflow speed and therefore
did not yield the equilibrium curves invoked in the LPP94 argument.

This issue was recently revisited by \citeauthor{CS02} (\citeyear{CS02},
hereafter CS02), who carried out a linear stability analysis on
an approximate equilibrium disk model. CS02 
derived a WKB dispersion relation and used it to infer
that magnetic wind-driving disks are unstable if the wind torque
is strong or (when the torque is weak) if the rotation is close
to Keplerian, but that magnetic diffusion stabilizes the disk if
this torque is small. As interpreted by LPP94 and
CS02, the instability reflects the sensitivity of the outflowing
mass flux to changes in the inclination of the magnetic field at
the disk surface: a perturbation increasing the inflow speed
causes the poloidal field to be bent closer to the disk surface,
giving rise to a higher mass loss and thereby a higher
angular momentum loss; this, in turn, allows more mass to flow
in, leading to a further increase in the inflow speed.

Although the equilibrium configuration examined by CS02 contains
more of the relevant physics than the model adopted by LPP94,
it is still unclear to what extent the approximations made
in that work have impacted the final result. For example, the
field line shape used in determining the location of the sonic
point and thence the mass outflow rate was not calculated
self-consistently. This is reflected in the fact that its adopted form
does not have an inflection point near the disk surface --- as it should
based on the WK93 results. Furthermore, the normalized
magnetic torque at the disk surface ($\tilde T_{\rm m 0}$ in the
notation of CS02) is equal to $|V_r|/2V_{\rm K}$ (where $V_r$ is
the mean radial
inflow speed and $V_{\rm K}$ is the Keplerian rotation speed),
which is typically $\ll 1$ even in disks where the magnetic
torque dominates the angular momentum removal (e.g., WK93). On
the other hand, the normalized diffusivity parameter ($\tilde
\eta$ in the notation of CS02) is typically ${\cal{O}}(1)$ in steady-state
disks where the diffusivity balances the inward advection of the
field by the accreting matter.\footnote{$\tilde \eta = 5$ in the
fiducial ambipolar diffusion-dominated disk solution presented by WK93.} Since,
according to the CS02 analysis, low-$\tilde T_{\rm m 0}$ and
high-$\tilde\eta$ disks are stabilized by magnetic diffusivity, it is
unclear whether one could conclude that real disk/wind systems
are unstable even if one were to accept the CS02 results.

Numerical simulations offer a potentially effective means of
resolving the stability question. However, global numerical
studies of a magnetic wind-driving disk incorporating a
magnetic diffusion mechanism that enables the system to reach a
steady state have only recently become feasible
\citep[e.g.,][]{CK02,CK04}. The results reported to
date are indicative of stability in that they show that the
disk/wind system evolves to a near-stationary state, but
they clearly do not yet settle the issue. In an
attempt to shed more light on the problem, this paper adopts the
equilibrium-curve approach first used by LPP94 but
applies it to a more comprehensive physical model. In
particular, using the disk/wind solutions of WK93
(\S~\ref{equilibrium}; see also \citeauthor{Li95}
\citeyear{Li95}, \citeyear{Li96}), it is
demonstrated (\S~\ref{stability}) that not all equilibrium
disk/wind structures are unstable. It is then argued (\S~\ref{conclusion})
that real disks likely correspond to stable configurations.

\section{Equilibrium Disk/Wind Model}
\label{equilibrium}

Although the WK93 model involves a number of simplifications, it
incorporates the full set of structure equations for the disk
and the wind and thus may be expected to capture the basic features
of an accretion flow in which the angular momentum is removed
through a centrifugally driven wind. The model assumes a
steady-state, axisymmetric
and geometrically thin disk that is in near-Keplerian rotation
around a central mass. The disk is threaded by a
large-scale, open magnetic field ${\bf B}$ that is symmetric about the
midplane (where $B_r=B_\phi=0$, using cylindrical coordinates
$r,\ \phi,\ z$). In anticipation of an application to
protostellar systems, the disk is taken to be weakly
ionized. In the solutions presented in this paper it is assumed
that the field is ``frozen'' into the
ion/electron component and influences the dominant neutral
component through ion--neutral collisions (resulting in
a relative drift --- referred to as {\em ambipolar diffusion}
--- between the ions and the neutrals). In these solutions it is
also assumed that the relative drift between ions and electrons
(associated with the Hall term in Ohm's law) and the radial
motion of the field lines both vanish.\footnote{WK93
demonstrated that solutions with a nonzero midplane radial ion
speed $V_{{\rm i},r0}$ do not differ qualitatively from those in
which the field lines do not move radially. The actual
situation depends on the global structure of the star/disk
system and may evolve with time. Note, however, that in the
disk formation model presented in \citet{KK02} the rotationally
supported, diffusive disk solution satisfies $V_{{\rm i}, r0}=0$.}

The disk structure is obtained by solving
the mass conservation relation as well as the radial, vertical,
and azimuthal components of the momentum equation for the
neutrals, assuming isothermality. The ion number density is
determined from ionization balance considerations. To further
simplify the problem, WK93 considered the molecular-gas regime
where the ion density is constant ($n_{\rm i} \approx 0.5\ {\rm
cm}^{-3}$); this should typically be applicable on protostellar
disk scales $\sim 10^2\ {\rm AU}$, where the disk column density
is sufficiently low
that cosmic rays can be assumed to penetrate and ionize the
entire vertical column. In view of the postulated low ionization the ion
momentum equation is approximated by equating the ion--neutral
collisional drag to the Lorentz force. The model also incorporates
the azimuthal and poloidal components of the induction equation
as well as the condition
${\bf \nabla \cdot B}=0$. WK93 restricted attention to a
narrow radial band centered on a given radius $r$ and matched
the disk solution to a
radially self-similar wind model, but very similar results are
obtained when a global self-similar model for both the disk and
the wind is considered \citep{Li96}.

For specified values of $V_{\rm K}$, isothermal sound speed
$C$, and midplane mass density $\rho_0$, the disk solution at
radius $r$ is determined by the values of the neutral--ion
coupling parameter $\eta$,\footnote{This parameter is the same as
$\tilde \eta$ in CS02.} defined as the ratio of the (Keplerian)
dynamical time to the neutral--ion momentum-exchange time
$\tau_{\rm ni}\propto 1/n_{\rm i}$, the magnetic field strength
parameter $a \equiv V_{\rm{A0}}/C=B_0/[(4\pi
\rho_0)^{1/2}C]$ (where $B_0$ is the field
amplitude at $z=0$), which gives the ratio of
the midplane Alfv\'en speed $V_{\rm A0}$ to $C$, and the
inflow-speed parameter $\epsilon \equiv - V_{r0}/C$. The solutions derived by
WK93 correspond to the ``strong coupling'' regime: $\eta > 1$ at
all heights. In this case the thermal pressure is not much
larger than the magnetic pressure at the midplane (i.e., $a$ is
not $\ll 1$), $B_r$ starts to increase already near $z=0$ and by
the time the disk surface
(subscript $s$) is reached it generally exceeds $|B_{\phi,s}|$
(leading to a strong magnetic squeezing of the disk), and the
midplane inflow speed is typically of the order of $C$ (i.e.,
$\epsilon \sim 1$). \citeauthor{Li96} (\citeyear{Li96}; see also
\citealt{W97}) showed that ``weakly coupled'' disk
configurations, in which $\eta < 1$
over the bulk of the vertical column, are also possible. In this
case $a$ can be $\ll 1$, $B_r$ only starts to grow well above the
midplane (after $\eta$ comes to exceed 1), with $B_{r,s}/|B_{\phi,s}|$
usually remaining $<1$ and magnetic squeezing being relatively
unimportant, and the mass-averaged inflow speed is typically
$\ll C$. \citet{KK02} argued that the
latter type of disk is likely to arise in the gravitational collapse of a
rotating, magnetic, molecular cloud core; however, in this paper
attention remains focused on the strong-coupling solutions
constructed by WK93.

In the WK93 derivation, the disk solution is matched to the
``cold'' radially self-similar centrifugal wind solution
of \citet{BP82} by imposing continuity of the mass flux and of
$B_r$ and $B_\phi$ at the nominal disk surface (the height
where the disk radial electric field component is equal to
$-V_{\rm K} B_z/c$, which is required to lie above a density
scale height in the disk). The details of the transition between the diffusive
disk interior and the highly conducting wind gas are not
considered in this model, and neither is the possibility of heat
injection in a disk corona.\footnote{The implications of gas
heating above the disk surface have been discussed
by \citet{CF00b} and \citet{FC04}; see also \citeauthor{OL98}
(\citeyear{OL98}, \citeyear{OL01}).} Although these issues are
relevant to the determination of the precise range of parameter
values that give rise to viable solutions (see \S~\ref{conclusion}), they
do not affect the basic properties of the equilibrium curves
obtained in the WK93 model. This is verified, inter alia, by the
fact that the extension of the WK93 disk solution above the nominal
disk surface has similar characteristics to those of
the base of the \citet{BP82} wind solution. Any model
refinements that focus on the transition region between the disk and
the wind should therefore have no significant effect on the stability
analysis carried out in \S~\ref{stability}.

The wind model incorporates
the Bernoulli and transfield (Grad-Shafranov) equations and its
solution yields the flow and magnetic field structures above the disk. The
wind model parameters are $\kappa$ --- the normalized mass-to-magnetic flux
ratio, $\lambda$ --- the normalized total (particle and magnetic) specific
angular momentum, and $B_{r,s}/B_z$; however, the condition that
the wind pass through the Alfv\'en critical point reduces the
number of independent parameters to two. Since $\kappa$ and $\lambda$
can be expressed in terms of the disk solution variables, the
Alfv\'en constraint serves to reduce the number of parameters in
the underlying disk model as well. The disk outflow is also required to
pass through a ``sonic'' critical point --- this yields
the ratio of the ``sonic''-point density to the midplane density, which 
enters into the relationship between the wind and disk
parameters.\footnote{WK93 approximated the outflow at this
point as still being largely diffusive, in which case
the relevant ``sonic'' speed is the thermal speed of sound. If,
however, the ions and neutrals are sufficiently well coupled
to effectively form a single fluid at this location, then
the critical point corresponds instead to the 
slow-magnetosonic point \citep[if the shape of the field lines
at the ``sonic'' point is assumed to be given; e.g.,][]{O97} or
the {\em modified} slow-magnetosonic
point \citep[if the transfield equation is also included in
the calculation ---  e.g.,][]{Li95,V00}.}

\begin{figure}
\resizebox{\hsize}{!}{\includegraphics[angle=-90]{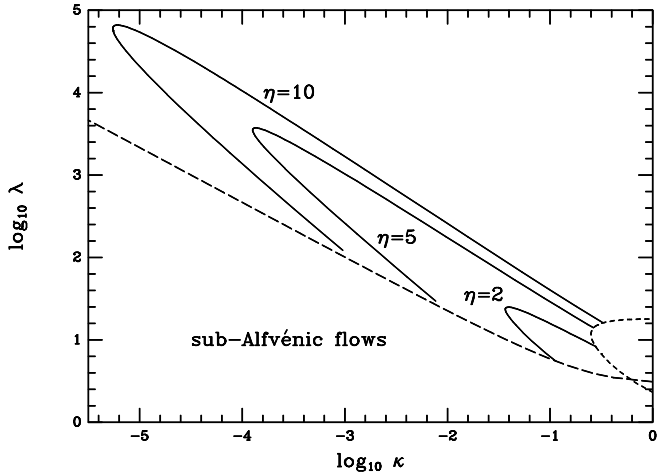}}
\caption[]{Mapping of the wind-driving disk solutions (for several values
of the neutral--ion coupling parameter $\eta$) onto the
self-similar wind solution space (defined by the values of the
mass-loading parameter $\kappa$ and angular-momentum parameter
$\lambda$). The field-strength parameter $a$ increases on moving
counterclockwise along a given curve.\label{fig1}}
\end{figure}

Figure \ref{fig1} (which reproduces Fig. 11 in WK93) shows the
mapping between the disk and wind solutions obtained in the WK93
model.\footnote{See Fig. 3 in \citet{Li95} for the analogous
result when the global self-similarity solution is extended to
also encompass the disk, with the field diffusion handled
using the generic Ohmic form and a turbulent-diffusivity
parameterization.} For given values of $\eta$ and $C$,
the disk solutions lie along a double-branched curve in the
$\kappa-\lambda$ wind parameter space. The
solutions along each curve are parameterized by $a$, which
increases from a value $\ll 1$ at the lower-right end of the
upper branch to a value $\sim 1$ at the lower-right end of
the lower branch. The lower branches of the solution curves end
on the long dashed curve, below which the outflows remain
sub-Alfv\'enic, whereas the upper branches end on the
short-dashed curve, to the right of which the surface layers of
the disk are super-Keplerian. As noted in WK93, $|B_{\phi,s}|/B_z$,
and hence the magnitude of the angular momentum that the outflow
must carry away, increases with decreasing $a$. Initially, as
$a$ decreases from $\sim 1$, $B_{r,s}/B_z$ (which scales
approximately as $1/a$) increases rapidly, and the corresponding
increase in the cylindrical radius of the Alfv\'en point (the
effective lever arm for the back torque exerted by the outflow
on the disk, which scales as $\lambda^{1/2}$) increases the
value of $\lambda$ and leads to a reduction in the ratio of the
mass outflow to the mass inflow rates (estimated as $\dot M_{\rm out}/\dot
M_{\rm in} \approx 1/[4(\lambda-1)]$). However, as $a$ continues
to decrease, the rate of increase of $B_{r,s}/B_z$ declines
while that of $|B_{\phi,s}|/B_z$
increases, and eventually the mass outflow rate must start to increase
(with $\lambda$ going down) to keep up with the angular momentum
removal requirements. The transition between these two modes of enhanced
angular momentum transport: predominantly by the lengthening of
the lever arm (on the lower branch) vs. mainly by a higher
mass-loss rate (on the upper branch) occurs at the turning point
of the solution curve.

\section{Stability Considerations}
\label{stability}

\begin{figure}
\epsscale{.96}
\plotone{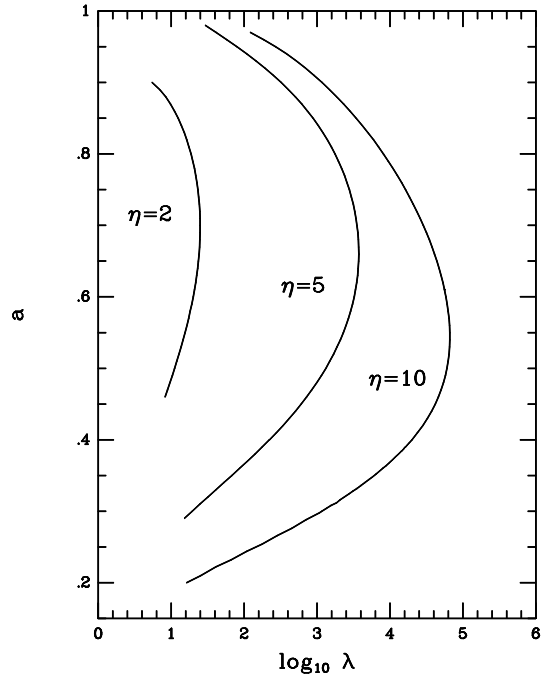}
\caption{Disk/wind solutions in the $\lambda-a$ plane.\label{fig2}}
\end{figure}

A turning point in the equilibrium curve typically
signals a change in the stability properties of the
corresponding solution. The usefulness of this approach
to the disk/wind solutions shown in Figure \ref{fig1} is perhaps best
demonstrated by replotting the curves in the $\lambda-a$
parameter space (Fig. \ref{fig2}, based on Fig. 12 in WK93). In
analogy with thermal stability analyses, one may think of the
parameter $a$ as representing temperature, the equilibrium value
of which is determined by the balance between heating and
cooling.\footnote{It is instructive in this regard to compare
the present discussion of the stability
properties of the WK93 magnetic disk model with Bell \& Lin's
(\citeyear{BL94}) analysis of the thermal ionization instability
in viscous protostellar accretion disks using the disks'
equilibrium vertical structure curves.}
Within the framework of the WK93 formulation, in which the
midplane density is time independent but the field lines can
potentially undergo a secular radial drift, the variation of $a \propto
B_0/\sqrt{\rho_0}$ is determined by
that of $B_z$ and is governed by the competition between inward
advection and outward diffusion (the analogs of cooling and
heating, respectively). Specifically,
\begin{equation}\label{Bz}
\frac{\partial B_z}{\partial t} =
-\frac{1}{r}\frac{\partial}{\partial r}(r V_{{\rm i},r}B_z)\ ,
\end{equation}
where $V_{{\rm i},r} = V_r + V_{{\rm D},r}$. For the solutions
shown in Figures \ref{fig1} and \ref{fig2}, the advection term
$V_r$ exactly balances the ion--neutral drift velocity $V_{{\rm
D},r}$ at $z=0$, so the midplane radial ion velocity vanishes
along the equilibrium curves.\footnote{The incorporation of a
possible imbalance between field advection and
diffusion into the WK93 model is analogous to the allowance for
a departure from vertical thermal balance in the \citet{BL94}
viscous disk solutions.}

In the case of a self-similar disk that
matches onto the \citet{BP82} wind solution,
$r V_{{\rm i},r}B_z$ scales as $r^{-3/4}$ \citep[e.g.,][]{Li96}
and hence the right-hand side of equation (\ref{Bz}), when evaluated at $z=0$,
has the same polarity (positive or negative) as $V_{{\rm i},r0}$. This 
implies that the value of $a$ will {\em decrease} with time if
advection dominates diffusion, which is consistent with the
physical picture of inward gas motion bending the field lines
and giving rise to a higher value of $B_{r,s}/B_z \propto 1/a$.
The polarity of $V_{{\rm i},r0}$ in turn depends on the relative
magnitudes of the advection ($<0$) and diffusion ($>0$) terms.
One can estimate the values of $V_r$ and $V_{{\rm D},r}$ from the angular
momentum and radial ion momentum equations, respectively, which
can be taken to have a steady-state form if
the timescale of interest is longer than the dynamical
(rotation) time and if the field is well coupled to the matter ($\eta >
1$). For a thin Keplerian disk with a symmetric field
configuration one obtains, to a good approximation,
\begin{equation}
V_{r0}\approx \left ( \frac{2 r}{h V_{\rm K}}\right ) \left
(\frac{B_z^2}{4\pi \rho_0}\right ) \left ( \frac{B_{\phi,s}}{B_z}\right
)
\label{Vr}
\end{equation}
and
\begin{equation}
V_{{\rm D},r0} \approx \left ( \frac{\tau_{\rm ni}}{h}\right )
\left ( \frac{B_z^2}{4\pi\rho_0}\right ) \left (
\frac{B_{r,s}}{B_z}\right )\ ,
\label{VDr}
\end{equation}
where $h$ is the disk scale height. As shown by WK93, $B_{r,s}$
generally exceeds $|B_{\phi,s}|$ for a strongly coupled disk, so
the condition of near-hydrostatic vertical equilibrium gives
$B_{r,s}/B_z \approx \sqrt{2}/a$. Furthermore,
$|B_{\phi,s}/B_z| = \kappa (\lambda -1)$ from the definition of
the \citet{BP82} self-similar wind
parameters. Hence equations (\ref{Vr}) and (\ref{VDr}) imply
\begin{equation}
\frac{|V_{r0}|}{V_{{\rm D},r0}}\approx \eta\,  a\,  \kappa\,
(\lambda -1)\ .
\label{ratio}
\end{equation}

Equation (\ref{ratio}) indicates that, if one moves to the right
of any given equilibrium curve in Figure \ref{fig2} by increasing
$\lambda$ (while keeping the other parameter values fixed), then the
value of $|V_{r0}|/V_{{\rm D},r0}$ will go up. In this
way we infer that advection dominates diffusion to the right of
the equilibrium curve (with the converse holding to the left of
the curve). It follows that, if $a$ is perturbed from its
equilibrium value, then, by equation (\ref{Bz}), the
perturbation will continue to grow if it starts on the lower
branch of the curve but will be counteracted if it originates on
the upper branch. From this we deduce that the lower branch of
an $\eta= {\rm const}$ equilibrium curve in Figure \ref{fig2} (or the
upper branch of the corresponding curve in Fig. \ref{fig1})
represents unstable solutions, whereas the upper branch in
Figure \ref{fig2} (or the lower branch in Fig. \ref{fig1}) is stable.

In the heuristic instability argument given by LPP94 and CS02,
a decrease in the field inclination to the disk surface
necessarily leads to an increase in the mass outflow rate from
the disk. However, as noted in \S~\ref{equilibrium}, the
equilibrium solutions of WK93 have revealed that a decrease in
the field inclination could also lead to a larger Alfv\'en radius
and that these two modes of angular momentum transport
act in competition with each other. As it turns out, the solution branch 
identified here as being unstable is in fact the one along which an
increase in the angular momentum transport is accomplished
through a higher mass outflow rate rather than through a
lengthening of the effective lever arm. Nevertheless, the inferred
existence of a stable branch indicates that a perturbation that
reduces the field inclination to the disk does not necessarily
trigger an instability. This may be understood from the fact
that an increase in $B_{r,s}/B_z$ also results in greater
field-line tension, which tends to oppose the inward poloidal
field bending. Whether a given solution branch is stable or not
is determined by the extent to which this (as well as any other relevant)
stabilizing effect can overcome the destabilizing influence of
increased angular momentum removal brought about by the
field-line bending.

\section{Conclusion}
\label{conclusion}

The stability properties of centrifugal wind-driving
accretion disks are inferred in this paper from a consideration
of the underlying equilibrium model. Specifically, the ambipolar
diffusion-dominated disk model of WK93 is shown to possess
equilibrium curves with two distinct branches, and it is argued
that the turning point between them marks a transition between
stable and unstable disk/wind configurations. Although this
remains to be verified, a growing perturbation on
the unstable branch may well correspond to the unstable mode identified by
CS02 and originally given a physical description by
LPP94. However, the existence of a generic stable branch implies that
such disks are {\em not} inherently unstable.

The nonlinear evolution of the instability considered by
LPP94 and CS02 has not yet been studied,\footnote{It is, however,
conceivable that this instability is related to the implosive soliton-like
accretion/outflow events discussed by \citet{LRN94}.}
so it is still unclear whether it results in a disruption of the disk. If this
were the case, then observed systems would naturally correspond
to the stable branch of the equilibrium solutions. In fact, in
some cases it may even be possible for a perturbed system on the
unstable solution branch to undergo a continuous parameter evolution 
into a stable configuration (see Figs. \ref{fig1} and \ref{fig2}).
One can, however, argue quite independently of the nonlinear outcome
of the instability that real wind-driving disks are likely
represented by {\em stable} solutions. One argument is
based on the results of \citet{F97} and \citet{CF00a}, who
refined the models used by WK93 and \citeauthor{Li95} (\citeyear{Li95},
\citeyear{Li96}) and concluded that the range of
viable solutions is reduced; in particular, they inferred that
all the allowed ``cold'' solutions lie on the equivalents of the {\em
lower} (i.e., stable) branches of the curves in
Figure \ref{fig1} (see Fig. 3 in \citealt{F97} and Fig. 5 in
\citealt{CF00a}). Another argument is based on the
likely formation mechanism of effectively steady disks in which the
ordered magnetic field originates on large spatial scales. In
this picture, which naturally applies to the formation of
protostellar disks from the collapse of rotating, magnetic,
molecular cloud cores \citep[e.g.,][]{KK02}, the field is
advected inward by the accreted matter but decouples from the
inflowing gas as the ratio of the magnetic diffusion time to the inflow time
decreases in the vicinity of the central mass. Under these
circumstances the field lines are not strongly bent and
$B_{r,s}/B_z$ is of the order of 1.\footnote{$B_{r,s}/B_z=4/3$
in the asymptotic ($r \rightarrow 0$) ambipolar
diffusion-dominated disk solution of \citet{KK02}.} Hence the
parameter $a$ remains $\la
1$ within the strongly coupled region of the accretion flow
($B_{r,s}/B_z$ would have been $\gg 1$ had $a$ been $\ll 1$) and
the corresponding equilibrium configurations lie on the {\em
upper} (i.e., stable) solution branches in Figure \ref{fig2}.

\lastpagefootnotes
The equilibrium-curve inferences presented in this paper need to
be corroborated by an explicit linear
stability analysis of the WK93 model as well as of more
elaborate equilibrium configurations. The arguments in support
of stable real disks are
consistent with the tendency of simulated disk/wind systems
to approach a near-stationary state (see \S~\ref{introduction}),
although it is worth noting that the equilibrium models and
disk/wind simulations presented to date have been limited to
axisymmetric structures. Since the disk and wind are
potentially susceptible also to various nonaxisymmetric modes
\citep[(e.g.,][]{KO00}, fully 3D simulations are required to
conclusively establish their stability
properties.\footnote{\citet{KLB03} and \citet{OCP03} inferred from
3D MHD simulations that steady-state centrifugally driven disk
outflows should be stable to nonaxisymmetric
perturbations. However, these calculations considered fixed
boundary conditions at the surface of the disk and did not
model the entire disk/wind system.} It is also worth keeping in mind that
the foregoing conclusions are based on a consideration of
strongly coupled disks, whereas (as noted in
\S~\ref{equilibrium}) the structure of protostellar
disks may at least in some cases correspond to the weakly
coupled regime discussed by \citet{Li96} and \citet{W97}. Since
in the latter type of disk the bulk of
the angular momentum is extracted from weakly coupled gas
that occupies most of the volume whereas the bending of field lines
only affects strongly coupled gas that resides in thin
surface layers, it may be expected that such configurations are
less susceptible to the field-bending instability
considered here than their strongly coupled
counterparts. However, a comprehensive study of the stability of
weakly coupled disks remains to be performed.

\acknowledgments
I am grateful to Z.-Y. Li and R. Rosner for helpful comments on
the manuscript and to N. Vlahakis for encouragement. This
research was supported in part by NASA Astrophysics Theory
Program grants NAG5-3687 and NNG04G178G.

\end{document}